\documentstyle[aps,preprint]{revtex}

\begin{document}


\title{Sr impurity effects on the  
magnetic correlations of La$_{2-x}$Sr$_x$CuO$_4$}


\author{R. J. Gooding and N. M. Salem}
\address{Dept. of Physics, Queen's University,\\ 
Kingston, Ontario, Canada.} 
\author{R. J. Birgeneau and F. C. Chou}
\address{Department of Physics and the Center for Materials Science
and Engineering,\\ Massachusetts Institute of Technology, Cambridge, 
Massachusetts, U.S.A.}

\date{\today}
\maketitle
\begin{abstract}

We examine the low--temperature magnetic properties of 
moderately doped ${\rm La_{2-{\em x}}Sr_{\em x}CuO_4}$, 
paying particular attention to the spin--glass phase and the 
commensurate--incommensurate transition as they are affected 
by Sr impurity disorder. A model of the carriers, believed to
be appropriate at low temperatures, is 
employed in the ${\rm CuO_2}$ planes that accounts for both the 
strong coupling of the hole's motion to the antiferromagnetically 
correlated spins and the pinning potential associated the 
${\rm Sr}$ impurities. This model has been 
shown to explain quantitatively several magnetic and transport 
features of
the antiferromagnetic region of the phase diagram. Preliminary 
indications that this
model can also explain the $x \geq 0.02$ region of the phase 
diagram follow
from its success in accounting for the doping and temperature 
variation of the spin correlation length. Here we further 
scrutinize this model by attempting
to explain various features of the spin texture of the spin--glass
phase. New measurements with travelling--solvent float zone grown 
crystals of the low--temperature susceptibility show an 
increase of an anomalously small Curie constant with doping. This 
behaviour is modelled in terms of
our numerical simulation results that find small clusters of 
antiferromagnetically 
aligned regions separated by disordered domain walls produced
by the impurities --- the domain walls lead to a percolating 
sequence of paths 
connecting the impurities. We predict that for this spin
morphology the Curie constant should scale as $1/(2~\xi(x,T=0)^2)$, 
$\xi$ being the spin correlation length,
a result that is quantitatively in agreement with experiment.
Also, we find that the magnetic correlations in the ground states 
produced by our 
simulations are commensurate in the spin--glass phase, consistent 
with experiment, and
that this behaviour will persist at higher temperatures
where the holes should
move along the domain walls being effectively expelled from the 
antiferromagnetically
correlated domains.  However, our results show that incommensurate 
correlations develop 
continuously around 5 \% doping with an incommensurate wave vector 
along ($(\pm 1,0), (0,\pm 1)$), consistent with recent measurements 
by Yamada, {\em et al.} 
At this doping level the domain walls and not the clusters start to 
become the dominant feature of the spin texture, and thus the Curie 
behaviour should disappear
in the incommensurate phase, again consistent with experiment.
Thus, we find that this model is capable of describing the 
low--temperature disorder--induced magnetic spin morphologies of the 
${\rm La_{2-{\em x}}Sr_{\em x}CuO_4}$ system in the low and
intermediate doping regions of the phase diagram.
Coupling this understanding of the magnetic correlations 
with the observed transport features,  it is clear that 
{\underline {impurity effects strongly influence the 
physics of LaSrCuO.}}


\end{abstract}
\newpage

\section{Introduction:}

The magnetic, electronic and transport behaviour of the 
${\rm La_{2-{\em x}}Sr_{\em x}CuO_4}$
high $T_c$ superconductor have been intensely studied, and one 
avenue of research that might lead to an understanding of the
anomalous normal--state properties, as well as possibly
the key to the superconducting instability, 
involves the examination of the evolution of the 
antiferromagnetic (AFM) insulator 
($0.0 \leq {\rm x} \lesssim 0.02$) to a spin--glass (SG) phase 
($0.02 \lesssim {\rm x} \lesssim 0.05$), and ultimately
to a superconductor with anomalous normal state properties 
($0.05 \lesssim {\rm x} \sim 0.2$).
Each of these doping ranges involves new and exciting physics with 
many 
mysteries to resolve \cite {rjbREVIEW,johnstonBOOK}, and in this 
paper we 
focus on the microscopic details of the spin texture of the SG phase.

Early experimental work demonstrated that the 
$0.02 \lesssim x \lesssim 0.05$ materials 
had a low--temperature SG phase, as evidenced in the freezing of the 
magnetic moment seen in $\mu$SR studies \cite {MUSR}. Shortly 
afterward it was noted that quasielastic neutron
scattering results for this phase \cite {haydenMFL} 
could be interpreted in terms of the freezing of microdomains 
roughly of a size of 20 \AA. 
Subsequent NQR experiments supported this conjecture by showing how
the La relaxations rates could be understood in terms of AFM domains
of small, finite--size magnetic domains \cite {choSG}. These 
interpretations made it quite natural to refer to this phase as a 
cluster SG.

More recent experiments have provided a much more complete
characterization of these compounds. Most importantly, the
measurements show evidence of
canonical SG behaviour \cite {chouSG}. Further,
samples in this doping range are found to have an anomalously 
small Curie constant as one approaches the SG transition from above 
\cite {chouSG}. Also, studies of the anomalous transport properties 
\cite {keimerCL}, {\em e.g.}, a conductivity that 
rises logarithmically with temperature over a small temperature 
range, and an isotropic, negative magnetoresistance 
\cite {preyerMR}, serve to 
emphasize the unusual behaviour of LaSrCuO in this doping regime.

In this paper we discuss how the spin texture of the SG phase
at a microscopic level \hfill\break 
(i) necessitates the identification of
this phase as a cluster SG, 
and (ii) explains how the small finite--size domains characterizing 
this spin texture are consistent with the anomalously small Curie 
constant found experimentally. 
To do this, we must begin with a model for
the effect of the dopant Sr impurities and the holes they induce into
the CuO$_2$ planes. However,
if a model is to be successful in describing the physics of this 
compound,
it must be able to describe the physics in {\em all} doping ranges,
not simply the SG regime.
We employ such a candidate model --- it was introduced previously
\cite {goodingSKYRM,goodingDJ} and is based on the strong
coupling of the hole's motion to the background antiferromagnetic 
Cu--spin correlations. To be specific, at low temperatures, when the
carriers are localized in the region of the Sr impurities, they generate
spin distortions that one can model using a spin--only Hamiltonian.
Our model is only strictly justified at low temperatures where the
carriers are localized near the Sr impurities --- however,
as has been argued elsewhere \cite {goodingDJ}, and as we elaborate on
below, we believe that the spin distortions produced by the holes that 
gradually become mobile as the temperature is raised will resemble those
generated by the localized carriers that we focus on in this paper,
and thus the qualitative aspects of the arguments presented here
should be applicable at higher temperatures.

This model accounts for the crucial and 
largely ignored role (in previous literature) of dopant 
{\em disorder}.  To be specific, when comparing its
predictive powers to the physics of the weakly--doped AFM phase 
this model
has been shown to be capable of describing (i) the traverse 
spin--freezing temperature {\em vs.} doping \cite {goodingTSF} 
observed by Chou {\em et al.} \cite {chouTSF}, and
(ii) provides a better fit to the conductivity {\em vs.} $T$
data than 3D variable--range hopping, as well as correctly
predicting the critical concentration $x_c \approx 0.02$ at 
which the metal non--metal transition takes place 
\cite {chenMNM,laiMNM}.
Further, when compared to the behaviour of the intermediate 
doping regime for the SG phase, preliminary work \cite{goodingDJ} 
showed that this model is able to track the behaviour of the doping
and temperature--dependent spin correlation length, 
as measured by Keimer {\em et al.} \cite
{keimerCL} and characterized by the empirical relationship 
\begin{equation}
\xi^{-1} (x,T) = \xi^{-1} (x,0) + \xi^{-1} (0,T)~~.
\label{eq:keimerCL}
\end{equation}

In this paper we continue the comparison of our candidate model
by studying its spin texture {\em vs.} doping in the SG regime.
We focus on understanding the spin texture at a microscopic level
and find that disorder induces meandering domain walls connecting
Sr sites; we argue that this is the correct way to think of the 
underlying spin texture of a cluster SG. A state qualitatively
similar to this but one produced by electronic phase separation in 
the absence of Sr impurity disorder was previously proposed by Emery
and Kivelson \cite {emeryPS}. In our model the possibility of 
electronic phase separation does not exist, and so electronic phase 
separation is not
necessarily the progenitor of the cluster spin SG. Instead,
we believe that the distortions generated by the holes and the Sr 
disorder are the key aspects of this problem.

A summary of our new theoretical and experimental results for this 
spin texture are as follows:

I~~ --- Clusters of antiferromagnetically correlated regions are 
formed, the boundaries of which are defined by the Sr impurities.
These domains have AFM order parameters which vary in direction,
and thus long--ranged AFM order in this doping regime is destroyed 
by the lack of orientational order of the moments of these clusters. 
The smallest collection of impurities
which forms a small cluster of AFM correlated spins 
is three non--collinear Sr impurities,
and from such three--impurity configurations it is established that 
domain walls separating the clusters are regions of disordered spins
which have a larger local ferromagnetic (FM) correlation than 
anywhere
else in the lattice. Also, since carriers are much more mobile in 
FM regions, it is reasonable to assume that this spin texture 
leads to meandering rivers of charge connecting the impurity 
sites --- namely, at the higher temperatures at which the
holes move the domain walls constrain and are populated 
with the mobile holes. 
This spin texture thus explains why the $x \lesssim 0.05$ material
does not develop incommensurate magnetic fluctuations: since
the holes are ``expelled" from the AFM clusters by their preference
for moving in a FM region, there is no impetus
for the local magnetic order to peak at any wave vector other than
${\bf Q} = (\pi,\pi)$.
We demonstrate this for the low--temperature spin texture by the 
explicit evaluation of $S({\rm Q})$.  

IIa --- The susceptibility experiments of Chou {\em et al.} on the 
SG phase of ${\rm La_{1.96}Sr_{0.04}CuO_4}$ between 20 K and 100 K 
displayed 
an anomalously small Curie constant \cite {chouSG}. We have grown
a new set of samples using the travelling--solvent, float--zone 
method,
for a variety of concentrations in the SG regime. We have determined
the Curie constants for all of these samples, and have determined 
how this constant changes with doping concentration in the SG 
phase.

IIb --- Modelling the topology of the SG phase in terms of small 
domains of 
AFM correlated spins separated by disordered domain walls,
and then treating the clusters as independent low--spin 
magnetic
units, we show that such a small moment follows immediately, and that
it increases with doping by one over the correlation length squared.
This result is in quantitative agreement
with experiment, and thus our comprehensive experimental study of 
this quantity serves as a critical test of the cluster SG 
phenomenology. 

III --- As mentioned previously, this model accounts for some 
observations in the AFM phase, and in this paper we show that
it also describes some results found in the SG phase.
If this model is an accurate representation
of the spin texture in all doping regimes it should also be able 
to explain
the appearance of incommensurate correlations for $x \approx 0.05$ 
\cite {endohIC}. Indeed, we find that this is the case, providing 
very strong 
support for this model. We associate this phenomenon with the 
degradation of 
the integrity of the domains of the cluster SG as doping levels are 
increased, and relate the delay in the appearance of the 
incommensurate 
correlations to impurity disorder effects. We propose that
our results are
suggestive that the mechanism behind the incommensurate phase 
follows from the spiral phase theory of Shraiman and Siggia 
\cite {shraimanSP}.

Our paper is organized as follows. Firstly, in \S II we discuss our 
model
of the spin distortions introduced by a single Sr impurity and a 
single hole,
and explain how this may be used to study the effects of a nonzero 
density
of holes all localized around Sr impurities. Then we present an
analysis of a small number of Sr impurities (three, in fact) that 
leads
to the simplest understanding of how domains of AFM correlated 
spins are
produced, and survey results from our extensive numerical simulations
that produce the ground states for our model Hamiltonian. In \S III 
we
present new experimental results for the susceptibilities of this
compound just above the SG transition temperature, and then show how
the above--mentioned small AFM domains adequately describe the 
observed
Curie constants. In \S IV we present a discussion of the 
commensurate--to--incommensurate transition, and relate the effects 
of spin distortions that we have included explicitly to theoretical
mechanisms for the spiral phase. Lastly, we summarize our results and
relate them to other outstanding problems in the study of weakly 
doped LaSrCuO. A brief, preliminary report on this work has already 
appeared elsewhere \cite {goodingHOUSTON}.

\newpage
\section{Model and Simulation Results:}

\subsection{Model Hamiltonian for a Single Hole around a Sr 
Impurity:}

It is well known that the near--neighbour Heisenberg Hamiltonian 
provides
a very accurate representation of the magnetic properties of the 2D 
CuO$_2$ planes
\cite {grevenCL,johnstonBOOK} apart from small anisotropies which 
we ignore in the present paper ({\em e.g.}, these anisotropies only
affect the correlation length when it becomes of the order of
80 lattice constants, a distance that greatly exceeds any length scale 
discussed in this paper). Thus, we represent the magnetic interactions 
between the Cu spins using 
\begin{equation}
 H_J= J~\sum_{<ij>}~{\bf S}_i \cdot {\bf S}_j
\label{eq:ham J}
\end{equation}
where $<ij>$ denote neighbouring Cu sites, and $J$ is the exchange 
constant, known to be of the order of 1550 K.

At low doping levels and at low temperatures, transport measurements 
\cite {chenMNM,keimerCL,laiMNM} suggest that the carriers move by hopping. 
Presumably their motion corresponds to the holes hopping from one 
impurity site to another, such as occurs in doped semiconductors 
before the metal--nonmetal
transition. Then, it is appropriate to consider the response of the
magnetic background to carriers localized in the vicinity of the Sr 
impurity ions.  A variety of previous theoretical 
studies \cite {goodingSKYRM,riceSKYRM,rabeSKYRM}
have all shown that the ground state for a single hole in such a 
situation corresponds to a doubly degenerate ground state for 
which the hole circulates
either clockwise or counterclockwise in the plaquette directly 
above or 
directly below the Sr impurity. The hole motion couples to the 
transverse fluctuations of the Cu spins and produces a spiraling 
twisting of the AFM 
order parameter \cite {goodingSKYRM,shraimanLONG}, a state that is 
topologically similar to the singly--charged skyrmion excitations 
of the 2D classical nonlinear sigma model \cite {belavinSKYRM}. 

One of us and Mailhot \cite {goodingDJ} proposed that a simple way 
of representing of the effects of this circulating hole motion 
on the AFM background was to integrate out the hole motion 
and replace it by a purely magnetic interaction. For (1,2,3,4) 
denoting the four spins in the plaquette bordering the Sr impurity, 
the interaction Hamiltonian for a single hole is given by
\begin{eqnarray}
 H_{int}=& - \frac{D}{S^4}
 \Big[ ({\bf S}_1\cdot {\bf S}_2\times {\bf S}_3)^2 
 +({\bf S}_2\cdot {\bf S}_3\times {\bf S}_4)^2 \cr
 &+({\bf S}_3\cdot {\bf S}_4\times {\bf S}_1)^{2}
 +({\bf S}_4\cdot {\bf S}_1\times {\bf S}_2)^{2} \Big ]~~.
\label{eq:ham int}
\end{eqnarray}
One may show that for classical spins, as long as $D/J > 2.2$ the 
ground state of $H_J + H_{int}$ has the same topology as that of 
the circulating hole ground state. 
The factor of $S^4$ ensures that this Hamiltonian scales as $S^2$, 
the same as the near--neighbour Heisenberg Hamiltonian (which 
facilitates the inclusion of quantum fluctuations in a 
straightforward manner), and the 
ratio $D/J \approx 3$ has been suggested \cite {goodingDJ} on the 
basis of (i) comparisons to Raman scattering results, and (ii) 
a numerical simulation
that has been shown to be consistent with Eq.~(\ref{eq:keimerCL}).
From now on we shall assume that this interaction Hamiltonian
represents the spin texture in the neighbourhood of any Sr 
impurity, and that we can represent the low--temperature multiply 
doped state with such
an interaction present at every plaquette bordering a Sr site.

\subsection{Impurity State Interactions:}

One may examine the interaction between pairs of these 
skyrmion--like impurity states
and explain the transverse spin--freezing temperature found 
experimentally
\cite {goodingTSF,chouTSF}. In this section we wish to consider the 
interactions
between more than two such impurity states, as this immediately 
leads one to an understanding of the possible origin of the cluster spin 
glass phase.

Around each impurity site the ground state is chiral, with the spin 
distortions
corresponding to either clockwise or counterclockwise circulations. 
On the basis of only this fact it is straightforward to predict the 
ground state for two Sr 
impurities: as shown in Fig.~1a, if one chooses opposite chiralities 
around the two Sr sites, then the ``currents" (either those of the 
motion of the hole,
or the concomitant spin currents --- see Ref. \cite {goodingDJ}) add 
constructively
in the region between the impurities. One may show, either 
numerically
or approximately analytically, that this is indeed the ground 
state. Now
consider three non--collinear Sr impurities. It is clear that 
there is no configuration for which the chiralities assigned to 
each impurity do not destructively interfere with one 
another --- three such impurity
states necessarily lead to a frustrated interaction between
the spin distortions created by each Sr impurity, as shown 
in Fig.~1b, and the question then arises: what is the ground state 
of this configuration of impurity sites? 

Figure 2 shows the ground state that we have found (our method for 
finding the ground state is detailed in the next subsection 
\cite {gooding3SR}) for three impurity sites 
placed in the positions of an isosceles triangle with the lengths 
of the sides roughly equivalent to the separations between randomly 
placed Sr impurities for a doping level in the middle of the SG 
phase. Figure 2a shows the actual
spin directions, whereas Fig. 2b shows the spins on all B 
sublattice sites flipped, as this substantially improves the 
visualization. The Hamiltonian is isotropic in spin space, and 
thus we have complete freedom to orient the 
spins in any direction --- we have chosen that the AFM order 
parameter would point directly out of
the page infinitely far from these impurities.

Figure 2b most clearly shows the very important result that the 
state that is produced 
by the frustrated current interactions is a small domain of AFM 
correlated spins
whose orientation is rotated with respect to that of the bulk AFM. 
For example, in the configuration
depicted in the figures the domain's AFM order parameter points 
toward the top of the page whereas the order parameter of the bulk 
AFM points out of the page.
Also, there are narrow domain walls separating this domain from the 
bulk of the lattice, although as expected theoretically 
\cite {goodingDJ,shraimanLONG} the spins distortions decay 
algebraically as one moves away from the Sr impurities.
With many Sr impurities distributed randomly above and below a 
plane, we thus expect to find many such domains, and we now 
analyze the results from our numerical
determination of the ground states showing that this is indeed an 
appropriate
way to think of this phase: these domains are the predominant 
components of the cluster SG spin texture.

\subsection{Ground State Spin Textures for the SG Regime:}

As explained above, the ground state for a random distribution of 
Sr impurities
is frustrated ({\em viz.}, the interactions between impurities
are frustrated). Thus, standard numerical approaches such as the 
conjugate--gradient
method usually find low--lying states, but not the ground state. 
We have numerically searched for the true ground states using
a technique first put forward to examine a random distribution of 
frustrating
FM bonds doped into a 2D AFM XY model on a square lattice 
\cite {grempelFM}, 
a technique
that from now on we shall refer to as the Grempel algorithm. This 
approach
was formulated after earlier work \cite {vannimenusFM} showed that
the lower energy states of this latter model could be thought 
of in terms of small domains of AFM correlated spins separated by 
domain walls, very similar to the spin texture that we are 
proposing for our perturbed Heisenberg model.

Given that the Grempel algorithm is helpful in illuminating the 
physics
of the cluster SG phase, we briefly explain its key features with 
regards
to the Sr impurity problem discussed in this paper. Firstly, for 
a given
lattice size and distribution of the Sr impurities we start by
assigning random directions to all spins.  Then, using the 
conjugate--gradient technique we ``relax" the system and 
find a state that is a local minimum on the SG energy 
surface. We repeat this for up to fifteen initial conditions 
(our results
did not change when we experimented and used more than fifteen).
Then, we applied the Grempel algorithm to find lower energy 
configurations: 
We begin with the lowest--energy relaxed state and first calculate 
the local field at each lattice site, the latter being defined by
\begin{equation}
{\bf H}_{i} = \sum_{j \epsilon <ij>} (-1)^j~{\bf S}_{j}~~.
\label{eq:local field}
\end{equation}
(Although ${\bf H}_{i}$ is not the exact local field for the spins
on the Sr--doped plaquettes, it nonetheless supplies an adequate
description of the disorder and frustration in the spin morphology.)
If these classical spins are assumed to be of a unit magnitude, then
the maximum magnitude of the local field is 4. Then we calculate 
the local field's magnitude for each spin, and if it is lower 
than some critical field $H^{*}$, we change its orientation to 
be any direction. For those sites 
which have a large local field, only a small random angle is 
used (chosen from no more than 5 \% of the 4 $\pi$ steradians on 
the unit sphere). This method ensures that large domains of AFM 
correlated spins have their integrity preserved
at this step, whereas the domain walls, the regions of more 
strongly disordered
spins (and thus smaller local fields), are allowed to more 
fully explore
the full spectrum of spin directions. Then, another local minimum is
found using the conjugate gradient method and the Grempel 
algorithm is
repeated with this new local minimum. This procedure allows the 
system to more
easily access its lower energy states. We have varied the critical 
field $H^{*}$ from $2$ up to $3.9$ after each application of the
conjugate gradient method, and find that after only 3 or 4 
applications
of the Grempel algorithm the (presumably) true ground state is
found. A full critique of the success of this method may be found
in Ref. \cite {grempelFM}.

We have found ground states (or at the very least, very low 
energy states)
for Sr doping concentrations corresponding to values of 
$x = 0.02, 0.035, 0.05, 0.065,$ and 0.08. 
Different lattice sizes, $L \times L$ with 
$L =$ 20, 30, 40, 50 and 60, 
and different distributions of Sr impurities, at least five for 
each $x$, are implemented as described in the Appendix.
In order to obtain better statistics we
also included $80 \times 80$ clusters for $x = 0.02$ and $x = 0.035$.
(We impose one restriction on each distribution of Sr 
impurities --- in an attempt to mimic the Coulomb repulsion 
between impurities that would
tend to force them apart during the crystal growth process, 
we do not allow for two near--neighbour plaquettes to both 
contain Sr impurities.) 

As a check that we are indeed finding ground states consistent with
the actual spin texture of LaSrCuO, from these ground states we 
obtained the following quantities:

\noindent
(i) The bulk value of the staggered magnetization. Assuming that the
finite--size scaling for the staggered magnetization is the same as 
it is for the undoped lattice {\em if} the system possesses 
long--ranged order \cite {FS},
{\em viz.} $M^{\dagger} (L) \sim M^{\dagger} (\infty) + O(1 / L)$,
we find that for $x = 0.035 \rightarrow 0.08$ 
long--ranged order is absent; for $x = 0.02$ finite--size effects
are too large for us to determine if long--ranged order 
is present.  Thus, it is unclear whether or not our $x = 0.02$ simulations 
are properly reflecting the absence of long--ranged AFM order found 
experimentally \cite {chouTSF} for $x \geq x_c = 0.0175 \pm 0.0025$. 

\noindent
(ii) The zero temperature correlation length ${\em vs.}$ doping. 
We have used
the finite--size scaling ideas that are covered in the Appendix to 
extract
this quantity from our ground states. Our results are shown in 
Fig. 3, along 
with the compilation of all two--axis experimental 
data contained in 
Ref. \cite {johnstonBOOK}.
Clearly, over the $0.035 \leq x \leq 0.08$ range of doping 
our numerically 
determined correlation lengths agree with experiment quite well. 
For $x = 0.02$ we find a correlation length that is significantly 
less than the
approximately 40 lattice constants found in the experiments of 
Keimer {em et al.}
\cite {keimerCL}. The possible reasons behind this failure are similar
to those discussed above:
For such a large correlation length we would
have to study systems that were of a linear dimension that were 
at the very minimum twice the correlation length.  Also, because of 
the low concentration, the average should be over more distributions of 
disorder to properly sample
what a low concentration bulk system would be like. Thus, for
the $x = 0.02$ simulations both finite--size
effects and a restricted averaging over disorder potentially limit
the reliability of our simulations and thus prevent
a proper comparison with experiment.
We have been able to fit our $\xi(x,T=0)$ data for 
$x \geq 0.035$ to the following empirical form (setting the 
lattice constant to be one from now on):
\begin{equation}
\xi(x,T=0)\ = {A\over {(x - x_c)^\eta}}
\label{eq:xi critical}
\end{equation}
We find that the best fit occurs for $A \sim 0.49, x_c = 0.0$ and 
$\eta \sim 0.98$,
and thus our data are reasonably consistent with the 
scaling $\xi(x,T=0) \propto 1/x$. This implies the vanishing of long--ranged
AFM order in the plane at T = 0 for any non--zero density of Sr impurities,
a result that may be understood in terms of the algebraic decay of the
spin distortions produced by a single hole localized near a Sr impurity.
The resulting curve is shown in Fig. 3, and clearly represents a 
good fit to both our numerical data and the data found in the 
study of the 3 and 4 \%
samples by Keimer {\em et al.} \cite {keimerCL}.


The $x = 0.02$ results described in the two points listed above 
suggest that, due to finite--size effects, we are unable to 
reliably simulate the $x = 0.02$ phase, and thus from now on 
we focus on samples with a minimum concentration of $x = 0.035$.

Encouraged by the $x \geq 0.035$ successes, we now describe the spin 
texture, at a microscopic level, that is associated with these 
ground states.
We proceed in a manner that most clearly displays the existence 
of the domains.
The first information that details the kind of inhomogeneous 
spin texture that exists in the SG phase is the local field 
defined in Eq.~(\ref{eq:local field}).
In Fig. 4a we show the local fields for a $x = 0.035$ 
distribution of Sr impurities
(this concentration is in the middle of the SG phase) on a 
$60 \times 60$ lattice.
The plaquettes having a Sr impurity above or below them are 
shown as open squares, and the size of the local field at the 
Cu sites is
indicated by the radius of a solid circle. For this figure all local fields less
than 99 \% of the maximum local field are not included. (Smaller 
values of this field do not affect the qualitative features
of this figure: the domain wall widths become smaller and the
continuity of the domain walls connecting neighbouring Sr sites
is not as clear, so here we use this relatively large local
field to make the domain walls' appearance as obvious as possible.)
This figure displays some evidence of 
clusters of strongly AFM correlated spins; clearly, some regions 
of spins are no longer strongly AFM correlated with 
other regions. 
However, what this figure most clearly shows is that there are 
channels of disordered 
spins (defined as having a reduced local field) that are set up 
connecting neighbouring 
Sr impurities --- these paths of disordered spins are the domain 
walls of the clusters.

More complete information concerning the actual spin texture that 
is found follows
from the kind of figure shown in Fig. 4b. This diagram corresponds 
to the
same ground state shown in Fig. 4a, and is constructed in the same 
way as 
Fig. 2 --- the projection of each spin onto the xy plane is shown, 
and every spin on a B sublattice site is flipped so that a 
completely ordered state would
appear to be a FM state with all spins pointing up out of the page. 
A careful
examination of this figure clearly shows the existence of the 
domains. For 
example, in the upper middle portion of this figure a region
bounded by six Sr impurities is seen to produce a 
domain of strongly AFM correlated spins. 
Further, one may characterize the sizes of the domains using
figures such as Fig. 4b --- we have done this for a number of
the ground states found in $60 \times 60$ lattices, and observed
that the average number of spins per domain varies as
$\xi(x,T=0)^2$ to within 10 \%.

Many other examples of these domains can be found, some being
three, four, $\dots$ sided domains. These assemblies are always 
associated with
the Sr impurities being at the edges of the clusters, and thus 
it is clear that 
these AFM correlated regions are a result of the dopant disorder. 
It is the
domain--like spin texture of this phase which motivates us to refer to it
as a cluster SG. 

One question that this spin texture answers is 
the elimination of long--ranged AFM order: We 
find that the orientations of the local AFM order parameters
of these clusters is random, and thus the disappearance of AFM order
may be thought of in terms of the lack of orientational
order of the local cluster moments. 
Thus, the long--ranged AFM order is replaced by the SG order. 

\subsection{Commensurability of the Spin Correlations in the 
SG Phase:}

One may calculate $S({\bf Q})$ for the ground states of the
$x = 0.02$ or $x = 0.035$ states, and it is always found that
the peak occurs at the commensurate AFM wave vector \hfill\break
${\bf Q} = (\pi,\pi)$.
(For higher doping concentrations incommensurate correlations 
develop, a subject that we shall discuss in \S IV.)
This is consistent with the experimental observation that the spin 
correlations remain
commensurate in the SG phase \cite {keimerCL,endohIC}. To see that
this may still be the case even if the holes are mobile simply 
note that holes
like to move in FM regions rather than AFM ones. Then, since the 
domain walls
are more disordered than the interior of the domains, they necessarily 
have a greater FM correlation present; thus, the holes will tend to
move along the domain walls.
Put another way, the holes are {\em expelled} from the interiors of the
domains, and only move along the domain walls.
(A complete theoretical discussion of the transport in the low $x$
compounds will be presented elsewhere \cite {laiMNM}.) 
This provides a plausible explanation of why the spin correlations remain 
commensurate both at low temperatures,
when the hole's motion is dominated by the Sr impurity potentials, 
or at higher temperatures, where the holes are mobile (and
a resistivity that varies linearly with $T$ is found). We note that
a similar explanation of the commensurability of the SG phase was 
pointed out by Emery and Kivelson \cite {emeryPS}.

\newpage
\section{Curie Constant:}

An important experimental result contained in the study of 
Ref. \cite {chouSG}
involves the behaviour of the susceptibility {\em vs.} temperature
just above the SG transition temperature, $T_G$. 
To be specific, $T_G \approx 10 $K
and it was found that from roughly 20 K to 100 K the susceptibility 
$\cal{X} (T)$
could be described via a Curie form with the Weiss constant 
equal to zero:
\begin{equation}
{\cal{X}} (T) = {\cal{X}}_0 + \frac{{\cal{C}}}{T}
\label{eq:Curie chi}
\end{equation}
For one well--studied $x = 0.04$ sample it was found that the Curie
constant $\cal{C}$ was very small, corresponding to
only $0.5 \%$ of the Cu spins present in the sample. Other samples
studied in the work associated with Ref. \cite {chouSG} were found
to have similar susceptibilities with Curie constants corresponding
to between $0.2 \%$ and $4 \%$ of the Cu spins present in the sample
\cite {chouUP}.  In this section we present new experimental results 
that detail the manner in which ${\cal{C}}$ varies with doping 
level, and then show how the spin texture described at the end 
of the previous section leads to a straightforward and
plausible explanation of this behaviour.

\subsection{New Experimental Results:}
 
Single crystals of ${\rm La_{2-{\em x}}Sr_{\em x}CuO_4}$ 
($x$ = 0.02, 0.025, 0.035, and 0.045) of masses around \hfill\break
2.0 $g$ have been grown using a crucible--free, contamination--free,
Travelling--Solvent Floating Zone (TSFZ) method \cite {TSFZ}.
The stoichiometric feed rods are prepared from the proportional 
amount of high
purity originals ($> 99.999 \%$) ${\rm La_2O_3,~SrCO_3}$ and CuO.  
The solvent rods are
prepared using CuO as self flux ($20 \%~{\rm La_2O_3,~80 \%~CuO}$).  
The optical floating--zone furnace uses four elliptical mirrors 
with halogen lamps at one focal point and the solvent zone is 
melted at the second focal point. The growth rate is about 
1 mm/hour, the rotation rate is 50 r.p.m. opposite, and 
the growth is done under air flow.

The as--grown single--crystal samples were
prepared using identical preparation techniques --- it was hoped that
this would produce a reliable trend of the Curie constant with doping
(see below). For the first three concentrations, this was the case.
Unfortunately, the $x = 0.045$ sample showed signs of 
superconductivity, and this concentration will only be addressed 
in the next section.  The as--grown crystals were subsequently 
exposed to additional treatments: oxygen annealing or vacuum 
annealing.  The oxygen annealing was performed at a temperature
of 600 $^\circ$C for 10 hours.  The vacuum annealing has been done at
850 $^\circ$C under $\approx 10^{-6}$ torr for one hour. The 
susceptibility results for these samples were very similar to 
those of
the as--grown samples, thus supporting our contention that these
are excellent single crystals with a high degree of Sr impurity
homogeneity.

The magnetization as a function of temperature has been measured
using a Quantum Design SQUID magnetometer.  The $M$ {\em vs.} $H$ data 
indicate that there are no ferromagnetic impurities present within 
our instrumental resolution.  The obtained $M$ {\em vs.} $T$ data were
replotted 
as ${\cal{X}}T$ {\em vs.} $T$ to extract the Curie constants from 
the region above $T_G$ in which the slope is a constant. 

The data for our samples are presented in Fig. 5; the 0.5 \% value 
for the Curie constant quoted for the $x = 0.04$ crystal studied in 
Ref. \cite {chouSG} is similar to that extracted from this figure.
The Curie constant is 
expressed as follows: For a given mass of LaSrCuO we calculate the 
Curie constant that would be expected if all of the Cu spins were 
isolated $S = 1/2$ spins with $g = 2$, and call this ${\cal{C}}_0$. 
Then, we express
the Curie constant found in the region above $T_G$ as a fraction of
${\cal{C}}_0$.  The errors in our data are roughly constant and equal 
to 10 \% of the measured values.

In the as--grown samples there is a clear indication of a linear 
variation with doping concentration, approximately given by 
${{\cal{C}}(x) / {\cal{C}}_0} = (0.1 \pm 0.02)~x~.$
For the O$_2$ annealed samples it could be argued that the Curie 
constant
grows quadratically with doping, whereas for the vacuum annealed 
samples it appears that the Curie constant is nearly independent 
of $x$ (we note, however, that the Oxygen depletion is expected to 
be largest for the higher Sr dopings for vacuum annealing, 
and thus our identification of hole doping
levels with $x$ is least accurate for this preparation technique).
Despite this variation, it is pleasing that in all cases 
the magnitude of the Curie
constant is roughly similar, and we now present a heuristic argument 
that accurately tracks this behaviour.

\subsection{Cluster SG Model of the Curie Constant:}

The results from the work of \S IIC suggested that the spin texture 
of the 
low--temperature moderately doped LaSrCuO system can be thought of 
in terms
of small clusters of AFM correlated spins separated by narrow 
domain walls
of disordered spins. The linear dimension of these domains tracks the 
zero--temperature
correlation length. Also, in the above section we showed how the 
Curie constant found in the temperature range just above $T_G$ 
varied 
with $x$, with an anomalously small number of effective Cu spins 
contributing to this constant.
In this section we show that these two results are compatible with 
one another,
and thus establish the consistency between phenomenologies 
promoting the 
existence of a cluster SG \cite {haydenMFL,choSG} and the 
experiments that 
showed that this material has a canonical SG phase \cite {chouSG}.

We describe the cluster SG spin arrangements in terms of domains
of spins, the sizes of the domains corresponding, on average, to the
zero--temperature correlation length, and  narrow domain walls
composed of strongly disordered spins. Due to the domain walls 
we make the
assumption that there is no interaction between the domains --- this
is consistent with the vanishing Weiss constant found experimentally.
Then, the effective Curie constant for this spin texture follows
immediately: As a simple picture of such spin configurations, let 
each domain be of a $L \times L$ geometry; further, take 
$L \sim \xi(x,T=0)$. Noting that
the quantum--mechanical ground state for such a cluster is a $S = 0$
singlet for $L$ even, and a $S = {1\over 2}$ doublet for $L$ odd, 
{\em and} that the gaps to the first excited states are 
large \cite {gapINFO}, as a first approximation we take the 
singlet clusters
to be magnetically inert (in the temperature interval over which
the susceptibility was measured). Accordingly, for a lattice with 
$N$ $Cu$ spins,
the susceptibility arises from $N / (2 L^2) $ spin--${1\over 2}$ 
magnetic units. Finally, for a cluster SG we predict
\begin{equation}
{{\cal{C}}\over{\cal{C}}_0} \sim {1\over 2 \xi^2}~~.
\label{eq:Curie Constant}
\end{equation}
The result of using Eq.~(\ref{eq:xi critical}) for the correlation 
length is also shown on Fig. 5, and displays the quantitative 
agreement between this admittedly highly simplified theory and 
experiment. Given that we
have ignored any contribution from the domain walls, it is not 
surprising that our value for the Curie constant is somewhat less 
than what is observed experimentally.
Further, Eqs.~(\ref{eq:xi critical},\ref{eq:Curie Constant}) 
together 
approximately yield ${{\cal{C}} / {\cal{C}}_0} \propto x^2$, 
consistent with the overall trend of an increasing Curie constant 
with $x$.  

\newpage
\section{Incommensurability {\lowercase {\em vs.}} Doping:}

In {\em all} ground states we have found some evidence for 
domains. These clusters are antiferromagnetically correlated, so 
it is not surprising that in the SG phase we found that the 
dominant magnetic correlations are commensurate, as we described 
in \S IID.
However, as $x$ increases we find that the areal fraction associated 
with the domain walls starts to dominate at the expense of the 
clusters, and thus the persistence of AFM correlations is no longer 
guaranteed. This is shown most clearly in Fig. 6 wherein we show 
a $x = 0.08$ result for the local fields
--- the $x = 0.035$ result for the same distribution of Sr 
impurities (at a lower
concentration --- see the Appendix) is shown in Fig. 4a. As is 
seen from a comparison of these two figures, the concept of 
domain walls is lost: only in a small number
of regions can one say that there are any clusters, and thus the 
concept of domain
walls separating the clusters is irrelevant. Instead, almost 
everywhere there is
a more homogeneously distorted spin texture, and effectively 
dopant disorder
is being averaged out. This is consistent with the experimental 
result \cite {endohIC} 
that for $x = 0.06$ one only observes the Curie behaviour 
over a much smaller
temperature range ($20 K \leq T \leq 50 K$), 
and the Curie constant that is found is much smaller, whereas
for higher $x$ no Curie--like behaviour is found at any temperature.
We now consider this behaviour in relation to the observed 
commensurate--incommensurate transition.

Recent experimental work by Yamada {\em at al.} \cite {endohIC} has
shown that the incommensurate (IC) correlations found in more 
strongly doped
LaSrCuO \cite {shiraneIC,cheongIC,masonIC,matsudaIC} 
develops continuously as the Sr doping
concentration is increased, with a critical concentration 
corresponding to \hfill\break $x_{IC} \approx 0.05$. It is very 
intriguing that
these correlations first develop at the $x$ value for which 
superconductivity
first appears, and thus makes the mechanism behind the appearance 
of the IC phase
an important question.  Here we show that our model, based on 
spin distortions produced by holes localized around Sr impurities
at low temperatures,
reproduces this commensurate--incommensurate transition. 
Most importantly, any model of mobile holes that induces spin 
distortions analogous to those discussed here may also produce
a similar IC phase (see below), and it can be 
argued \cite {goodingDJ}
that what we are analyzing are ``snapshots" of the kinds of 
effects produced by mobile holes generating dipolar backflow 
spin distortions.

We have only calculated the ground state spin textures for this 
system.
Thus, we cannot directly compare our results to the measured
dynamic structure factors ${\cal{X}}^{''}~({\bf Q},\omega)$ found
experimentally. Instead, we examine $S({\bf Q})$, and look for signs
of incommensurate correlations in the snapshots of the ground
state spin texture. 

We find that at $x = 0.05$ two of the five distributions of Sr 
impurities
generate IC correlations. For $x = 0.065$ three of the five 
distributions
generate IC correlations, whereas for $x=0.08$ all five of the 
distributions
of Sr impurities generated IC correlations. From these numbers 
we conclude
that the C--IC is seemingly continuous and happens close 
to $x \sim 0.06$. This
number is very close to that found experimentally.
Also, in {\em both} experiment and our simulations the 
correlations develop along $(\pm 1,0)$ or $(0,\pm 1)$. 
This agreement between our work and experiment is striking, 
and suggests that
mechanisms which produce the IC phase by spin distortions, such 
as the spiral
phase mechanism of Shraiman and Siggia \cite {shraimanSP}, 
should be re--examined 
with an eye to understanding the stability of this phase in 
the presence of strong Sr impurity disorder (albeit with a bias
towards maximum spin distortions around the Sr impurities).

\newpage
\section{Summary:}

We have considered a simple model of the effect of holes at low
temperatures that are localized 
around the Sr impurities in LaSrCuO and have found evidence for 
clusters of AFM correlated
spins in the SG phase --- we believe that this spin texture provides 
ample reason to refer to this state as a cluster SG. That this
two--fold degenerate chiral ground state leads to a SG phase
lends support to the prescient speculations of Villain \cite {villain}.
We briefly note that we have also considered another 
related model of spin distortions produced by localized O holes, 
the so--called Aharony model, and we still find strong evidence for such 
clusters in this model of Sr impurity disorder. 
Thus, our result is seemingly model independent 
{\em providing} that the Sr disorder is included explicitly.

We have examined in greater detail, using new TSFZ grown 
crystals, the anomalously
small Curie constant that was measured in Ref. \cite {chouSG}, 
and have
characterized both its magnitude and the manner in which this constant 
changes with doping.  We have thereby provided a quantitatively 
accurate model of this 
Curie constant by utilizing the idea that the clusters behave 
like independent
low--spin quantum units.

We have also shown that the C--IC transition in our model 
is continuous, and
occurs at roughly the same $x$ as is found experimentally. 
We thus attribute
the critical value of $x_{IC}$ to Sr impurity disorder. 
Also, our model
induces incommensurability in the same axial directions 
($(\pm 1,0), (0,\pm 1)$)
as is found experimentally. Given that the spin distortions 
produced by
the partially localized holes studied in this paper are similar 
to the
dipolar backflow spin distortions induced by mobile holes in 
the theories
of Shraiman and Siggia, it is possible that a model of such 
holes including the
effects of Sr disorder for $T \gg T_G$ 
will provide an accurate representation of the
physics of the intermediately doped LaSrCuO system studied here.
We note that previous theoretical claims of an incipient phase separation
of the spiral phase must be reinvestigated using a model
that includes explicitly the Sr disorder; it
is likely that disorder will completely suppress this instability.
We will report on the results of related studies in a future publication.

In none of the theoretical work presented here have we included 
the quantum fluctuations
of the Cu spins, instead relying on a classical model of the 
spins. Support
for this idea follows in part from the success of the 
renormalized classical
model in describing the undoped phase. However, we are now 
utilizing inhomogeneous,
linearized, spin--wave theory, including the modified 
spin--wave theory constraint
to enforce the fact that there is no long--ranged order in the 
SG phase, to assess the importance of quantum fluctuations to these 
results. Further,
upon completion of these results, we should better be able to 
test if the dynamical
fluctuations of such a cluster SG phase are responsible for the 
unusual scaling
of ${\cal{X}}^"_T(\omega)$ found by 
Keimer {\em et al.}\cite {keimerCL}.

We have also not included the XY anisotropy, a component that could 
possibly affect our results. However, the topology of a cluster SG 
was first found in a 2D XY model \cite {vannimenusFM,grempelFM}, and 
thus it seems highly unlikely that this small interaction would 
qualitatively affect our conclusions, especially given the
small AFM correlation length in the SG phase. Other magnetic anisotropies 
would serve to sharpen the domain walls separating the clusters, 
and thus could only serve to strengthen our arguments.

Lastly, we note that the ideas expressed in this paper are not
necessarily limited to the Sr--doped 214 compound. Recent
work by Budnick, {\em et al.} \cite {budnick123}, has shown that
Sr--doped 123 (with the Sr substituting for the Y ions) produces
a similar SG phase, and by symmetry we expect similar impurity 
ground states in this material. Also, Li--doped 214,
studied originally by Endoh {\em et al.} \cite {endohLi}, and more
recently examined in Ref. \cite {fiskLi}, 
potentially has chiral impurity states, as recently
discussed by Haas, {\em et al.} \cite {haasLi}. So, it is possible
that similar physics might also be found in this compound.

\newpage
\acknowledgments

We wish to thank David Johnston for providing us with a collection 
of the 
measured correlation lengths, and for other valuable comments. 
We also wish 
to thank Barry Wells, Martin Greven, Marc Kastner, Pakwo Leung 
and Ferdinando Borsa 
for helpful comments.  This work was initiated while one of us 
(RJG) was visiting 
MIT, and he wishes to thank them for their warm hospitality. 
This work was 
supported by the NSF, grant number 93--15715 and by the NSERC of Canada.

\newpage
\appendix{Appendix:}

One of the outstanding problems of theoretical solid--state physics 
involves the determination of accurate finite--size scaling forms for 
disordered systems.
We have attempted to circumvent this problem in the following manner:
We have performed our numerical simulations of this spatially 
disordered
system by imagining that we have ${\cal{N}}$ large samples of the
crystal, all having the same average density of Sr impurities above
and below a CuO$_2$ plane. Then, for one of the ${\cal{N}}$ large 
samples, choose some square subsystem of
the sample of dimension $L_1 \times L_1$, and find its ground state
spin texture. Then, as schematically shown in Fig. 7, 
we have studied increasingly larger subsystems, namely
$L_2 \times L_2$ and then $L_3 \times L_3$, {\em etc.}, with
$L_1 < L_2 < L_3 \dots$ that include exactly the same distribution 
of impurities in the smaller subsystems as are present in the 
larger subsystems. 
Then, as one increases the density of Sr impurities we add more 
impurities
to the initial distribution to raise $x$ to the desired value. 
From such sets of data we have found that we can best obtain good
statistics for each sample, and then we perform an average over
all ${\cal{N}}$ samples. Further, the manner in which we increase the
density of impurities leads to smoother variations of quantities 
with $x$.
We repeat this process for a number of differing initial, low density
distributions of impurities, and we find that quantities like the 
correlation length are very well converged with only five samples 
per $x$.

\begin{figure}
\caption{(a) A schematic of the interaction between competing 
circulations for two 
Sr impurity sites. When the chiralities of these states are 
opposite, one finds that 
the currents add in a constructive fashion. This is the ground 
state for any two
impurities. (b) If one adds a third non--collinear Sr impurity 
to the previous situation, 
neither a positive nor negative circulation for the third impurity 
leads to only 
constructive current interactions, and thus this situation is 
frustrated.}
\label{fig:currents}
\end{figure}

\begin{figure}
\caption{(a) The distribution of spin directions for the ground state
surrounding 3 non--collinear Sr impurities (shown by solid circles). 
What is shown is the projection of the spin's direction onto the 
plane of the 
paper (to aid in the understanding of these figures, a spin lying 
entirely
in the page pointing to the top of the page is shown at the right). 
Then, the blue spins
represent those spins pointing up out of the paper, while the red 
spins are those that are pointing down below the paper. The net 
orientation of
the spins is chosen to make the AFM order parameter (a quantity that
vanishes in the thermodynamic limit) for the bulk of the lattice
point up out of the page.
(b) The same as in (a) with the B sublattice spins flipped 
(such that an AFM would appear as a FM). This aids in the 
visualization of the spin directions, and more directly shows how 
the orientation of the AFM order
parameter in the domain formed by the 3 Sr impurities 
(lying approximately 
in the plane of the page pointing to the top of the page) 
is rotated with
respect to the direction of the AFM order parameter 
(pointing directly up out of the page) in the bulk of the lattice.}
\label{fig:3_Sr_sites}
\end{figure}

\begin{figure}
\caption{The correlation length determined from our ground states 
as a function 
of $x$ at $T = 0$ (the open stars) in comparison to all of the 
two--axis neutron 
scattering data collected since 1989 (from the collection of data 
in Ref. [2])
with the reference numbers included inside of angular brackets. 
The solid line is the fit of our data to the expression given in 
Eq. (5) with $A = 0.49$ and $\beta = 0.98$, and shows how well 
this empirical
form fits both our numerical results and the 3 and 4 \% samples 
studied in Ref. [7].}
\label{fig:CLs}
\end{figure}

\begin{figure}
\caption{These pictures explain the spin texture of a $x = 0.035$ 
system for one
particular distribution of Sr impurities on a $60 \times 60$ 
lattice. The plaquettes containing Sr impurities are shown as 
squares. In (a) the local fields,
as defined in Eq. (4), are shown. Clusters of various sizes and 
shapes are shown,
but the most striking feature of this picture is the presence of 
percolating paths
of disordered spins ({\em viz.}, spins that are not 
antiferromagnetically 
correlated) that connect the Sr impurities. In (b) the projection 
of the spins in the middle of this lattice (a $30 \times 30$ region) 
onto the plane of the page is shown, similar to the 
construction in Fig. 2a.}
\label{fig:x=0.035 spin texture}
\end{figure}

\begin{figure}
\caption{The variation of the effective Curie constant, 
expressed as a percentage
of the contribution that would arise from independent 
$S = 1/2$ ions with $g = 2$ at each Cu site, for several different 
LaSrCuO samples {\em vs.} $x$. 
The theoretical prediction contained in Eq. (7), using the
correlation length expression given in Eq. (5), is also shown.}
\label{fig:C vs x}
\end{figure}

\begin{figure}
\caption{The same as Fig. 4a for $x = 0.08$. There are fewer 
clusters present
in this ground state, and instead this spin texture is found 
to be much more 
homogeneous. Such a state has a Curie constant that is much 
smaller than that
predicted by Eq. (7) due to the absence of clusters. This 
behaviour is
consistent with experiment.}
\label{fig:x=0.08 local fields}
\end{figure}

\begin{figure}
\caption{This figure shows smaller lattices embedded in larger 
lattices. The
idea behind this construction is that what we want to study is small
chunks of an infinite system, and see how one approaches the 
infinite system by taking larger and larger sections. Since we are
considering a disordered system, translational periodicity is 
broken and
we must make sure that the larger clusters contain the same 
distribution of impurities as the smaller clusters. 
Thus, in practice we place 
one distribution of Sr impurities into the largest 
(say, $L_3 \times L_3$) 
cluster, and then examine the properties of the $L_1 \times L_1$, 
$L_2 \times L_2$, and then $L_3 \times L_3$ clusters, {\em etc.}, 
to best 
represent the manner in which the system approaches the bulk limit.}
\label{fig:disordered_FSS}
\end{figure}


\begin{references}
\bibitem{gapINFO}
For example, this gap is equal to 0.28 $J$ for a 32--site square 
lattice, 
$J$ ($\approx$ 1500 K for $LaSrCuO$) being the AFM exchange constant.
\end{references}
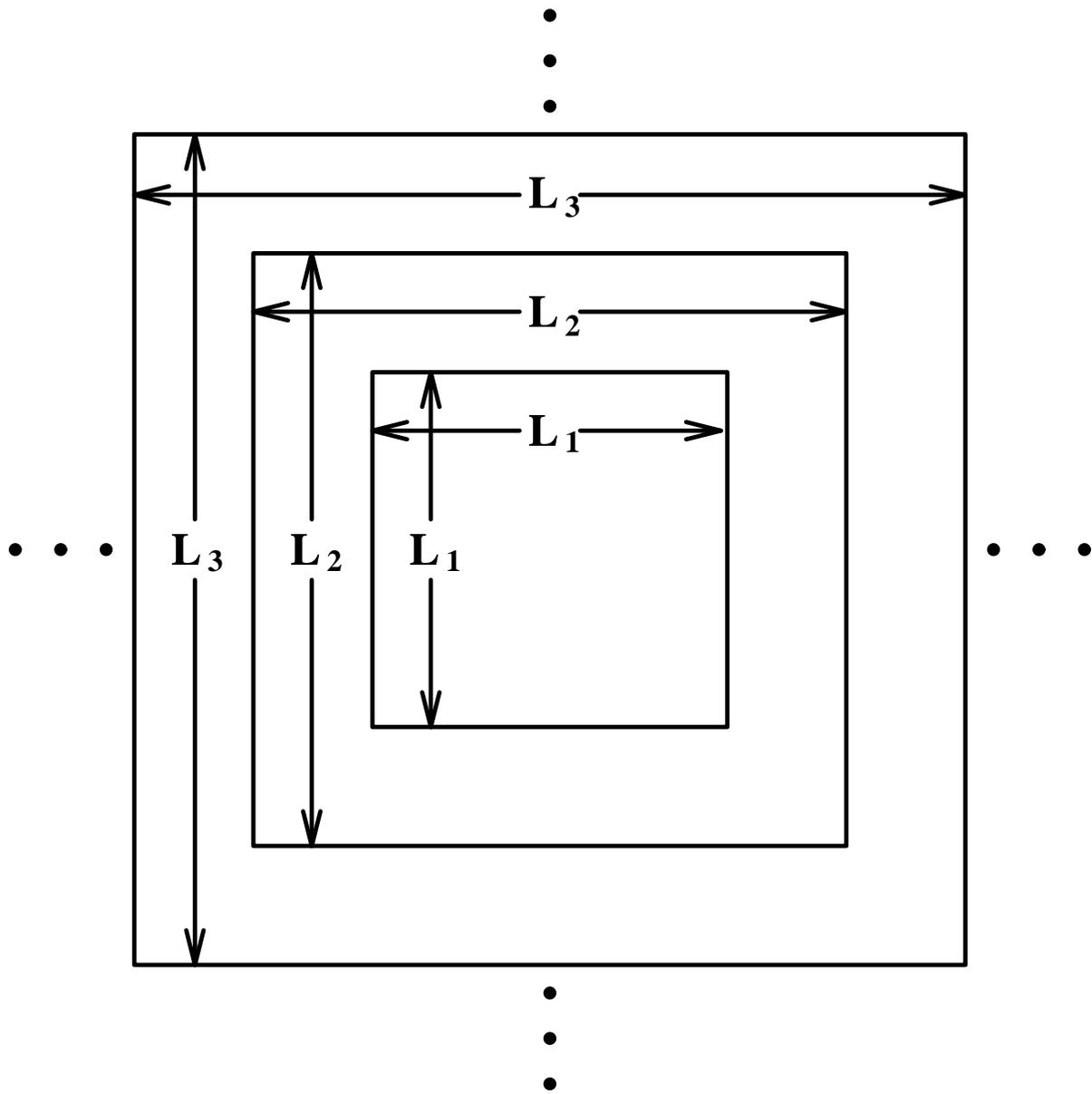
\end{document}